\documentclass[aps,preprint,superscriptaddress,]{revtex4}
\usepackage{amsfonts}
\usepackage{amsmath}
\usepackage{amssymb}
\usepackage{graphicx}
\usepackage{float}
\usepackage{subfigure}
\usepackage{epsfig,graphics,subfigure,psfrag,amsmath,amssymb}
\usepackage[justification=centering]{caption}
\begin{document}

\title{Effect of acceleration on information scrambling}
\author{Xi Ming}
\thanks{Corresponding author. Electronic address: mingxi@wipm.ac.cn}
\affiliation{Innovation Academy for Precision Measurement Science and Technology, Chinese Academy of Sciences, Wuhan, 430071, China}
\affiliation{University of Chinese Academy of Sciences, Beijing, 100049, China}

\begin{abstract}
The research subjects of information scrambling and the Unruh (anti-Unruh) effect are closely associated with black hole physics. 
We study the impact of acceleration on information scrambling under the Unruh (anti-Unruh) effect for two types of tripartite entangled states, namely the GHZ and W states. 
Our findings indicate that the anti-Unruh effect can result in stronger information scrambling, as measured by tripartite mutual information (TMI). 
Additionally, we show that the W state is more stable than the GHZ state under the influence of uniformly accelerated motion. 
Lastly, we extend our analysis to $N$-partite entangled states and product states.
\end{abstract}

\maketitle

\section{Introduction}

Information scrambling is a captivating interdisciplinary subject that merges quantum information and black hole physics~\cite{hayden2007black, shenker2014black, maldacena2016bound, sekino2008fast, landsman2019verified}. 
Information scrambling describes the dispersion of information from local degrees of freedom into the many-body degrees of freedom within a system. 
Generally, black holes are regarded as the fastest quantum scramblers in nature~\cite{sekino2008fast}. 
Recent research highlights the use of information scrambling as a potent tool for characterizing black hole chaos\cite{hayden2007black, shenker2014black, maldacena2016bound}. 
Beyond black hole physics, scrambling is extensively studied in quantum information~\cite{mi2021information, harris2022benchmarking}, quantum many-body dynamics~\cite{sahu2019scrambling, altman2018many, iyoda2018scrambling}, and quantum machine learning~\cite{shen2020information, holmes2021barren}. 
Multiple methods have been proposed to quantify quantum information scrambling, including out-of-time-order correlators~\cite{larkin1969quasiclassical}, the Loschmidt echo~\cite{yan2020information, sanchez2020perturbation}, and tripartite mutual information (TMI)~\cite{hosur2016chaos, sunderhauf2019quantum}. 
It is important to note that TMI is an operator-independent quantity and becomes negative when scrambling occurs throughout the system. Furthermore, smaller TMI values signify stronger information scrambling.     

Another intriguing phenomenon intimately connected to black hole research is the Unruh effect~\cite{unruh1976notes}. 
Inspired by the Hawking radiation resulting from black hole evaporation~\cite{hawking1974black, hawking1975particle}, Unruh postulated that a uniformly accelerated observer in the vacuum field would detect a thermal bath with temperature $T=a /2 \pi$, now known as the Unruh effect. 
It is evident that the temperature $T$ of this thermal bath is proportional to the acceleration $a$, paralleling the temperature of Hawking radiation~\cite{hawking1974black, hawking1975particle}. 
In addition to Hawking radiation, the Unruh effect contributes significantly to understanding other theories, such as cosmological horizons~\cite{gibbons1977cosmological} and the thermodynamic interpretation of gravity~\cite{jacobson1995thermodynamics, verlinde2011origin}. 
The Unruh-DeWitt model, which comprises a two-level point monopole coupled with a massive or massless scalar field along its trajectory~\cite{dewitt1979quantum}, is the most prominent proposal for investigating field-detector interactions. 
Under the Unruh effect, the transition probability of the detector increases with acceleration. 
Interestingly, recent research has found that the transition probability of an Unruh-DeWitt detector decreases as its acceleration increases under certain conditions.  
This new phenomenon is distinct from the traditional Unruh effect and is called the anti-Unruh effect~\cite{brenna2016anti, garay2016thermalization}. 

Previous research has explored the influence of both the Unruh effect and anti-Unruh effect on various quantum resources~\cite{fuentes2005alice, alsing2006entanglement, martin2009fermionic, martin2010unveiling, wang2011multipartite, bruschi2012particle, shamirzaie2012tripartite, richter2015degradation, dai2015killing, li2018would, pan2020influence, pan2021anti, wu2022genuine}. 
In this paper, we focus on the impact of acceleration on information scrambling, as quantified by TMI. 
We adopt the Unruh-DeWitt model and assume that three observers, Alice, Bob, and Charlie, each possess identical detectors that interact with their respective massless scalar fields. 
Here, we consider three detectors with different initial entangled states, including GHZ and W states. 
We examine the evolution of TMI with acceleration when one, two, or all three observers move at the same rate. 
Computational results reveal that the anti-Unruh effect can induce stronger information scrambling than the Unruh effect across the entire system. 
Besides, we observe weaker information scrambling in the W state, indicating that it is more robust than the GHZ state when subjected to uniformly accelerated motion. 
We also calculate the TMI for detectors initially in $N$-partite GHZ and product states.

This paper is organized as follows. In section ~\ref{sec:setup}, we briefly review the Unruh-Dewitt model and introduce the TMI for quantifying information scrambling. In section ~\ref{sec:three}, we investigate the evolution of information scrambling for several different situations. In section ~\ref{sec:n}, we extend the discussion to the case of multiple detectors. Finally, We conclude our work in section ~\ref{sec:conclusion}. 

\section{Setup}
\label{sec:setup}

\subsection{The Unruh-DeWitt model}
Firstly, we provide a brief overview of the Unruh-DeWitt model in this section. 
Consider a ($1+1$)-dimensional single-mode massless scalar field $\phi$ interacting with a two-level quantum system, with ground state $|g\rangle$ and excited state $|e\rangle$ separated by an energy gap $\Omega$. 
The interaction Hamiltonian for this model is~\cite{dewitt1979quantum}
\begin{equation}
	H_I=\lambda \chi(\tau / \sigma) \mu(\tau) \phi[x(\tau), t(\tau)],
\end{equation}
where $\lambda$ denotes the coupling constant, $\tau$ represents the detector's proper time along its trajectory $x(\tau)$ and $t(\tau)$, $\chi(\tau / \sigma)$ is the switching function controlling the interaction time via parameter $\sigma$, and $\mu(\tau)$ is the detector's monopole moment. 
We assume weak coupling with $\lambda = 0.1$, Gaussian switching function $\chi(\tau / \sigma)=\mathrm{e}^{-\tau^2 / 2 \sigma^2}$, and $t(\tau)=a^{-1} \sinh (a \tau)$ and $x(\tau)=a^{-1}[\cosh (a \tau)-1]$. 
For the weak coupling, the time evolution operation can be perturbatively expanded as
\begin{equation}
\begin{aligned}
	U&=I-i \int d \tau H_I(\tau)+O\left(\lambda^2\right) \\
        &= -i \lambda \sum_m\left(I_{+, m} a_m^{+} \sigma^{+}+I_{-, m} a_m^{+} \sigma^{-}+\text {H.c. }\right)+\mathcal{O}\left(\lambda^2\right),
 \end{aligned}
\end{equation}
where we treat the spacetime as a cylinder with spatial circumference $L=200$, $m$ indicates the discrete mode of the scalar field with periodic boundary conditions $(k=2 \pi m/L)$, $a_m (a_m^{+})$ represents the annihilation (creation) operator in mode $m$ of the scalar field, and $I_{\pm, m}$ is given by
\begin{equation}
	I_{\pm, m}=\int_{-\infty}^{\infty} \frac{1}{\sqrt{4 \pi|m|}} \chi(\tau / \sigma) \mathrm{e}^{\pm i \Omega \tau+\frac{2 \pi \mathrm{i}}{L}[|m| t(\tau)-m x(\tau)]} \mathrm{d} \tau.
\end{equation}
Within the first-order approximation, this evolution follows that~\cite{brenna2016anti,li2018would, wu2022genuine}
\begin{equation}
	\begin{aligned}
		&U|g\rangle|0\rangle =C_0|g\rangle|0\rangle-i C_0\eta_0|e\rangle\left|1\right\rangle,\\
		&U|e\rangle|0\rangle =C_1|e\rangle|0\rangle+i C_1\eta_1|g\rangle\left|1\right\rangle,
	\end{aligned}
\end{equation}
where $C_{0,1}=1 / \sqrt{1+\eta_{0, 1}^2}$ are normalization factors, and $\eta_{0, 1}=\lambda \sum\limits_{m \neq 0} I_{\pm, m}$ are connected with the excitation and deexcitation probability of the detector. 
In summary, we consider the massless scalar field case and remove the zero mode in a periodic cavity. 
It is worth mentioning that the validity of the above conditions has been demonstrated in ~\cite{garay2016thermalization}.

Assume an Unruh-DeWitt detector coupled to the scalar field is in the following state initially
\begin{equation}
\left|\psi\right\rangle =(\alpha \left|g\right\rangle + \beta \left|e\right\rangle)  \left|0\right\rangle.
\end{equation}
After accelerating, the above state will evolve into
\begin{equation}
  \begin{aligned}
     |\psi^{'}\rangle &= \alpha C_0 (|g\rangle|0\rangle-i \eta_0|e\rangle\left|1\right\rangle) + \beta C_1(|e\rangle|0\rangle+i \eta_1|g\rangle\left|1\right\rangle) \\ 
    &= \alpha |g\rangle \left|\psi_{0}\right\rangle+ \beta |e\rangle\left|\psi_{1}\right\rangle, 
  \end{aligned}
\end{equation}
where $\left|\psi_{0}\right\rangle=C_0|0\rangle+i(\beta/\alpha)C_1 \eta_1 \left|1\right\rangle $ and $\left|\psi_{1}\right\rangle=C_1|0\rangle - i(\alpha/\beta)C_0 \eta_0 \left|1\right\rangle $. 
The decoherence process can be quantified by the decoherence factor $D = |\langle \psi_{0} | \psi_{1} \rangle|$~\cite{bassi2013models}. 
The value of $D$ ranges between 0 and 1, with larger values indicating stronger coherence. 
Furthermore, the decoherence factor $D$ can also characterize the Unruh (anti-Unruh) phenomena~\cite{li2018would}. 
In general, the Unruh effect weakens coherence, while the anti-Unruh effect enhances it. 
As shown in Fig. 1, coherence tends to increase with acceleration when $\Omega = 0.1$ and decrease with acceleration when $\Omega = 2$. 
Thus, we can infer that $\Omega = 0.1$ and $2$ represent the anti-Unruh effect and the Unruh effect, respectively, under suitable conditions. 
This result is consistent with previous research that utilized transition probability to determine the occurrence of the Unruh effect or anti-Unruh effect~\cite{brenna2016anti}.
\begin{figure}
\begin{minipage}[t]{0.5\textwidth}
\centering
\includegraphics[width=0.9\textwidth]{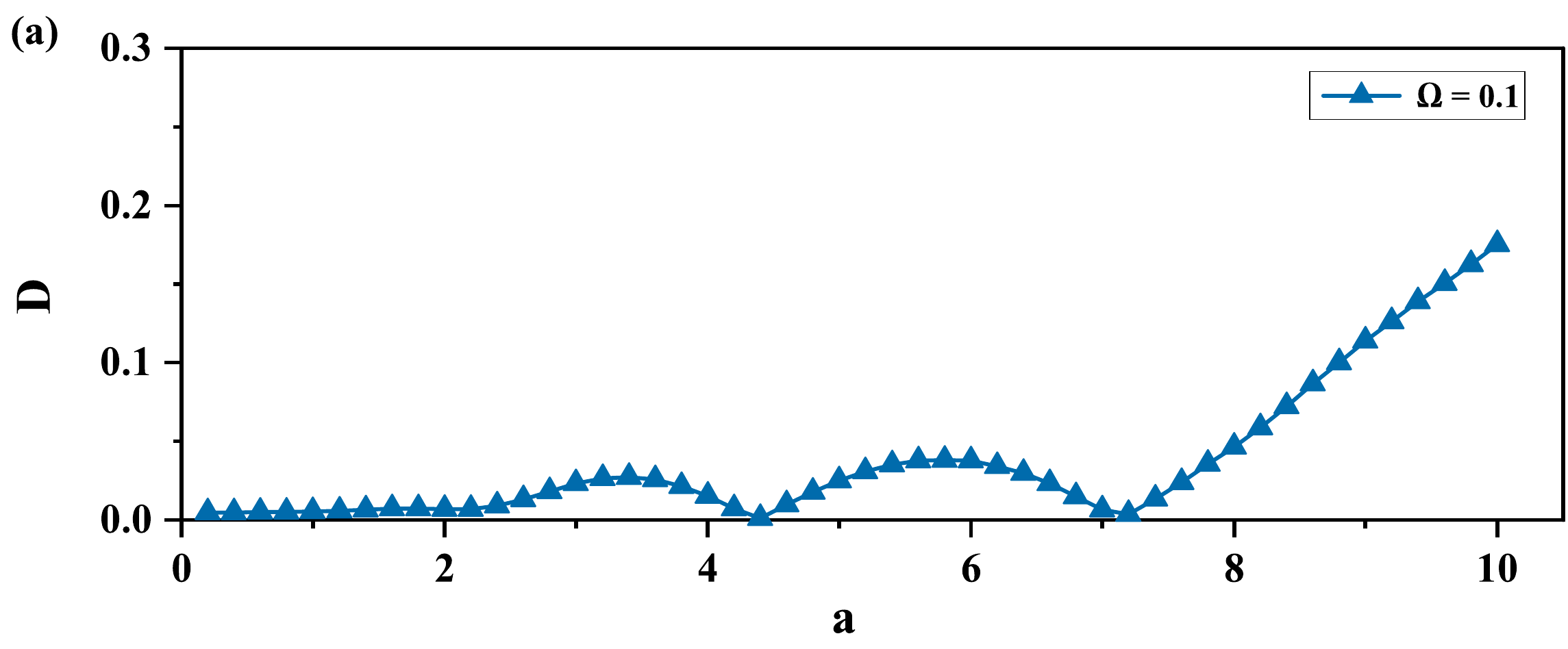}
\label{fig:1a}
\end{minipage}%
\begin{minipage}[t]{0.5\textwidth}
\centering
\includegraphics[width=0.9\textwidth]{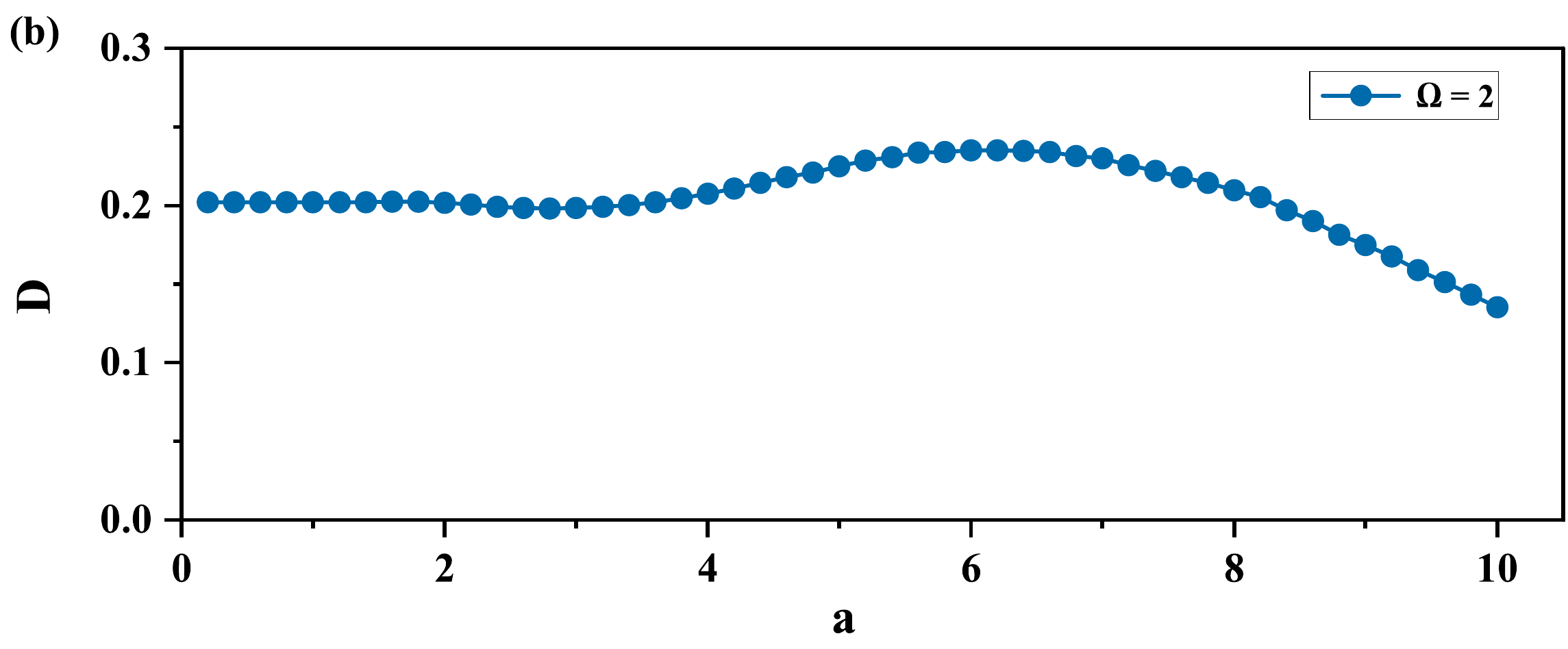}
\label{fig:1b}
\end{minipage}
\caption{The decoherence factor $D$ as functions of the acceleration $a$ with $\lambda=0.1$, $\sigma=0.4$, and $L=200$: (a) $\Omega=0.1$; (b) $\Omega=2$.}
\end{figure}

\subsection{Tripartite mutual information (TMI)}

For a tripartite system that contains three regions $A$, $B$, and $C$, we can introduce the von Neumann entanglement entropy, which is defined as 
\begin{equation}
    S_{A}= -\operatorname{tr} \rho_{A} \ln \rho_{A},
\end{equation}
where the reduced density matrix is $\rho_{A}=\operatorname{tr}_{BC} \rho_{A B C}$, and similar calculations are for the $\rho_{B}$, $\rho_{C}$, $\rho_{A B}$, $\rho_{A C}$, and $\rho_{B C}$.
Then, we can calculate the TMI by the following form~\cite{hosur2016chaos, sunderhauf2019quantum}, 
\begin{equation}
\begin{aligned}
    I_{3} =& I(A,B)+I(A,C)-I(A,BC) \\
    =& S(\rho_{A})+S(\rho_{B})+S(\rho_{C})+S(\rho_{A B C})-S(\rho_{A B})-S(\rho_{A C})-S(\rho_{B C}),
\end{aligned}
\end{equation}
where the bipartite mutual information $I(A, B) = S(\rho_{A}) + S(\rho_{B}) - S(\rho_{A B})$ measures the total amount of correlations between subsystems A and B, and analogous definitions for $I(A, C)$, $I(B, C)$, and $I(A, BC)$. 
Due to the subadditivity of von Neumann entanglement entropy, the bipartite mutual information must be non-negative. 
However, the TMI can be negative when the information of subsystem A stored in composite BC is larger than the total amount of information of subsystem A stored in subsystem B and subsystem C individually, i.e., $I(A, BC) > I(A, B) + I(A, C)$~\cite{iyoda2018scrambling}. 
This indicates that scrambling occurs and local information spreads from the subsystem throughout the entire system. 
Furthermore, $I_3 = 0$ if the tripartite system is a pure state or if subsystems A, B, and C are not related to each other, meaning the entire system is in a separable state. 
In this paper, we will utilize the TMI to quantify information scrambling.

\section{The case of three detectors}
\label{sec:three}

In this study, we examine three same Unruh-DeWitt detectors, $A$, $B$, and $C$, which are distant from each other and separately held by Alice, Bob, and Charlie. 
We assume the initial states of the detectors include two types of entangled states: the GHZ state and the W state, as follows:
\begin{equation}
  \begin{aligned}
    &|\psi\rangle_{A B C}^{GHZ}=\frac{1}{\sqrt{2}}\left(\left|g_A g_B g_C\right\rangle+\left|e_A e_B e_C\right\rangle\right)\left|0_A 0_B 0_C\right\rangle, \\
    &|\psi\rangle_{A B C}^W=\frac{1}{\sqrt{3}}\left(\left|g_A g_B e_C\right\rangle+\left|g_A e_B g_C\right\rangle+\left|e_A g_B g_C\right\rangle\right)\left|0_A 0_B 0_C\right\rangle,
  \end{aligned}
\end{equation}
where we treat the vacuum as being in a product state. 
In subsequent sections, we will investigate three different scenarios in which one, two, or all three detectors are accelerating.
\subsection{Alice in acceleration}
Initially, we consider only Alice moving with constant acceleration. 
Applying the transformation from Eq. (4), the initial GHZ state transforms into the following form,
\begin{equation}
\begin{aligned}
    |\psi\rangle_{A' B C}^{GHZ}=& U_A |\psi\rangle_{A B C}^{GHZ}\\
    =& \frac{1}{\sqrt{2}} ( C_0\left|g_A g_B g_C 0_A 0_B 0_C\right\rangle-i C_0 \eta_0\left|e_A g_B g_C 1_A 0_B 0_C\right\rangle + C_1 \left|e_A e_B e_C 0_A 0_B 0_C\right\rangle\\
    &+i C_1 \eta_1\left|g_A e_B e_C 1_A 0_B 0_C\right\rangle ).
\end{aligned}
\end{equation}
By tracing out the scalar field degrees of freedom, 
we can get the reduced density matrix $\rho_{A' B C}$ of three detectors, 
\[
\begin{pmatrix}
\frac{1}{2}C_0^2 & 0 & 0 & 0 & 0 &0 & 0 & \frac{1}{2} C_0 C_1^*\\
0 & 0 & 0 & 0 & 0 & 0 & 0 & 0\\
0 & 0 & 0 & 0 & 0 & 0 & 0 & 0\\
0 & 0 & 0 & \frac{1}{2}C_1^2 \eta_1^2 & -\frac{1}{2}C_0^* C_1 \eta_0^* \eta_1 & 0 & 0 & 0\\
0 & 0 & 0 & -\frac{1}{2}C_0 C_1^* \eta_0 \eta_1^* & \frac{1}{2}C_0^2 \eta_0^2 & 0 & 0 & 0\\ 
0 & 0 & 0 & 0 & 0 & 0 & 0 & 0\\ 
0 & 0 & 0 & 0 & 0 & 0 & 0 & 0\\
\frac{1}{2} C_0^* C_1& 0 & 0 & 0 & 0 &0 & 0 & \frac{1}{2}C_1^2
\end{pmatrix}
.\]
Further, we can obtain the reduced density matrixes $\rho_{A'}$, $\rho_{B}$, $\rho_{C}$, $\rho_{A' B}$, $\rho_{A' C}$, and $\rho_{B C}$, and the TMI is given by
\begin{equation}
\begin{aligned}
I_3=&C_0^2\log{(\frac{1}{2}C_0^2)} +C_1^2\log{(\frac{1}{2}C_1^2)}+(1 - C_0^2)\log{[\frac{1}{2}(1-C_0^2)]}+(1 - 
    C_1^2)\log{[\frac{1}{2}(1 - C_1^2)]}\\
    &-\frac{1}{2}(2 - C_0^2 - C_1^2)\log{[\frac{1}{2}(2 - C_0^2 - C_1^2)]} -
    \frac{1}{2}(1 + C_0^2 - C_1^2)\log{[\frac{1}{2}(1 + C_0^2 - C_1^2)]} \\
    &- \frac{1}{2}(1 - C_0^2 + C_1^2)\log{[\frac{1}{2}(1 - C_0^2 + C_1^2)]} - 
    \frac{1}{2}(C_0^2 + C_1^2)\log{[\frac{1}{2}(C_0^2 + C_1^2)]}+\log{2}
\end{aligned}
\end{equation}
Similarly, the initial W state will become 
\begin{equation}
\begin{aligned}
    \left|\psi\right\rangle_{A' B C}^{W}=& U_A |\psi\rangle_{A B C}^{W}\\
    = &\frac{1}{\sqrt{3}} ( C_0\left|g_A g_B e_C 0_A 0_B 0_C\right\rangle-i C_0 \eta_0\left|e_A g_B e_C 1_A 0_B 0_C\right\rangle+C_0\left|g_A e_B g_C 0_A 0_B 0_C\right\rangle\\
    &-i C_0 \eta_0\left|e_A e_B g_C 1_A 0_B 0_C\right\rangle + C_1 \left|e_A g_B g_C 0_A 0_B 0_C\right\rangle+i C_1 \eta_1\left|g_A g_B g_C 1_A 0_B 0_C\right\rangle ).
\end{aligned}
\end{equation}
And we can derive the reduced density matrix as 
\[
\begin{pmatrix}
\frac{1}{3}C_1^2 \eta_1^2  & 0 & 0 & 0 & 0 & -\frac{1}{3} C_0^* C_1 \eta_0^* \eta_1 & -\frac{1}{3} C_0^* C_1 \eta_0^* \eta_1 & 0\\
0 & \frac{1}{3}C_0^2 & \frac{1}{3}C_0^2 & 0 & \frac{1}{3}C_0 C_1^* & 0 & 0 & 0\\
0 & \frac{1}{3}C_0^2 & \frac{1}{3}C_0^2 & 0 & \frac{1}{3}C_0 C_1^* & 0 & 0 & 0\\
0 & 0 & 0 & 0 & 0 & 0 & 0 & 0\\
0 & \frac{1}{3}C_0^* C_1 & \frac{1}{3}C_0^* C_1 & 0 & \frac{1}{3}C_1^2 & 0 & 0 & 0\\
-\frac{1}{3} C_0 C_1^* \eta_0 \eta_1^*  & 0 & 0 & 0 & 0 & \frac{1}{3}C_0^2 \eta_0^2 & \frac{1}{3}C_0^2 \eta_0^2 & 0\\ 
-\frac{1}{3} C_0 C_1^* \eta_0 \eta_1^*  & 0 & 0 & 0 & 0 & \frac{1}{3}C_0^2 \eta_0^2 & \frac{1}{3}C_0^2 \eta_0^2 & 0\\ 
0 & 0 & 0 & 0 & 0 & 0 & 0 & 0
\end{pmatrix}
.\]
Eventually, if the initial state is W entangled state, the TMI can be calculated by 
\begin{equation}
\begin{aligned}
 I_3= &- \frac{1}{3}(3 - 2C_0^2 - C_1^2)\log{[\frac{1}{3}(3 - 2C_0^2 - C_1^2)]} 
    - \frac{1}{3}(1 + 2C_0^2 - C_1^2)\log{[\frac{1}{3}(1 + 2C_0^2 - C_1^2)]} \\
    &- \frac{1}{3}(2 - 2C_0^2 + C_1^2)\log{[\frac{1}{3}(2 - 2C_0^2 + C_1^2)]} - 
    \frac{1}{3}(2C_0^2 + C_1^2)\log{[\frac{1}{3}(2C_0^2 + C_1^2)]} \\
    &+\frac{1}{3}(2 - C_1^2 - \delta_1)\log{[\frac{1}{6}(2 - C_1^2 - \delta_1)]} +\frac{1}{3}(2 - C_1^2 + \delta_1)\log{[\frac{1}{6}(2 - C_1^2 + \delta_1)]} \\
    &+\frac{1}{3}(1 + C_1^2 - \delta_2)\log{[\frac{1}{6}(1 + C_1^2 - \delta_2)]} +\frac{1}{3}(1 + C_1^2 + \delta_2)\log{[\frac{1}{6}(1 + C_1^2 + \delta_2)]} \\
    &+\frac{2}{3}\log{\frac{3}{2}} + \frac{1}{3} \log{3} 
\end{aligned}
\end{equation}
where $\delta_1=\sqrt{4 - 4C_0^2 + 4C_0^4 - 4C_1^2 + C_1^4}$ and $\delta_2= \sqrt{1 - 4C_0^2 + 4C_0^4 + 2C_1^2 + C_1^4}$.

Next, we plot the information scrambling quantified by TMI$(I_3)$ as functions of acceleration $a$ for both $\Omega=2$ (Unruh effect) and $\Omega=0.1$ (anti-Unruh effect) in Fig. 2. 
The upper two plots (blue lines) correspond to the initial GHZ state, while the lower two plots (red lines) represent the initial W state.
Noticeably, the TMI value is approximately -1.00 in Fig. 2(a), which is less than the value of around -0.8 in Fig. 2(b). 
This relationship is also observed in Fig. 2(c) and Fig. 2(d). 
The results suggest that the anti-Unruh effect leads to stronger information scrambling. 
Moreover, the TMI value in Fig. 2(a) is smaller than that in Fig. 2(c), and the TMI value in Fig. 2(b) is also smaller than that in Fig. 2(d). 
This demonstrates that detectors in the W initial state are more robust than those in the GHZ state under the influence of acceleration.
\begin{figure*}
\begin{minipage}[t]{0.5\textwidth}
\centering
\includegraphics[width=0.9\textwidth]{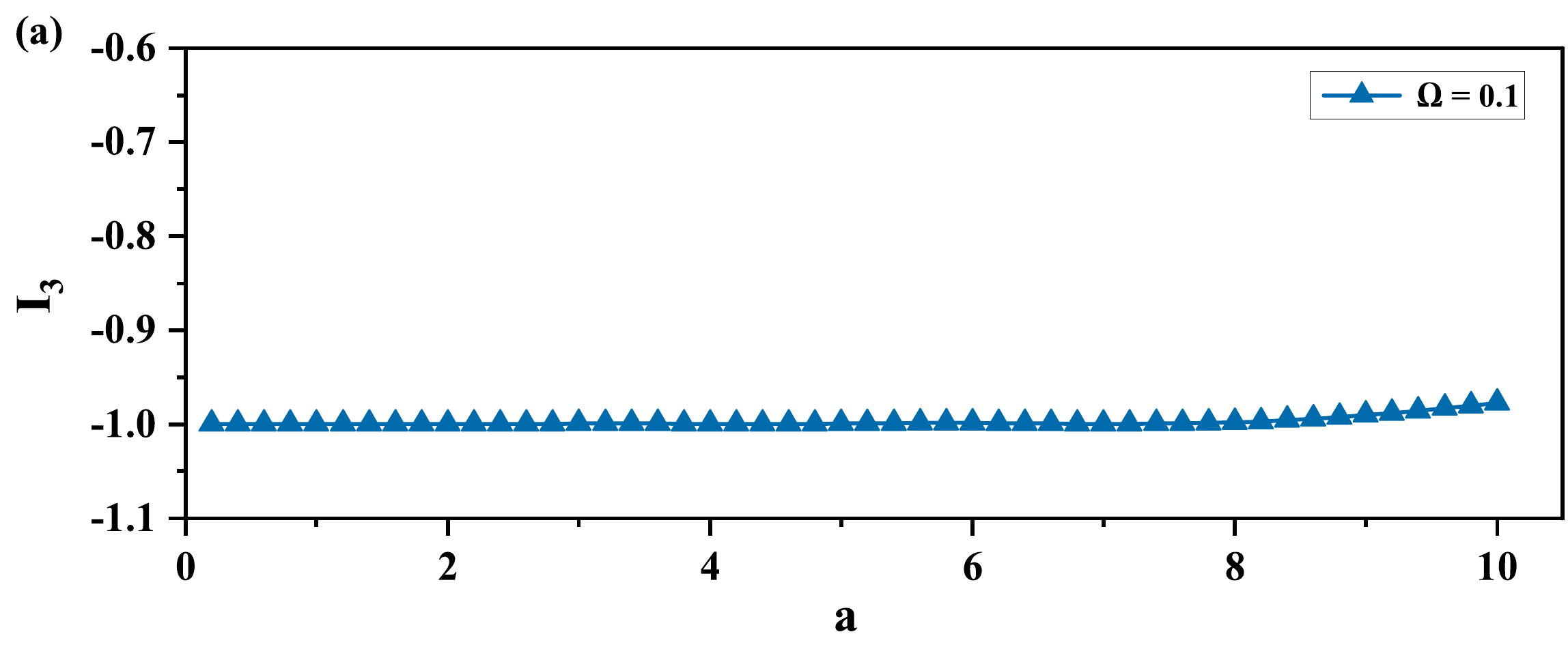}
\label{fig:2a}
\end{minipage}%
\begin{minipage}[t]{0.5\textwidth}
\centering
\includegraphics[width=0.9\textwidth]{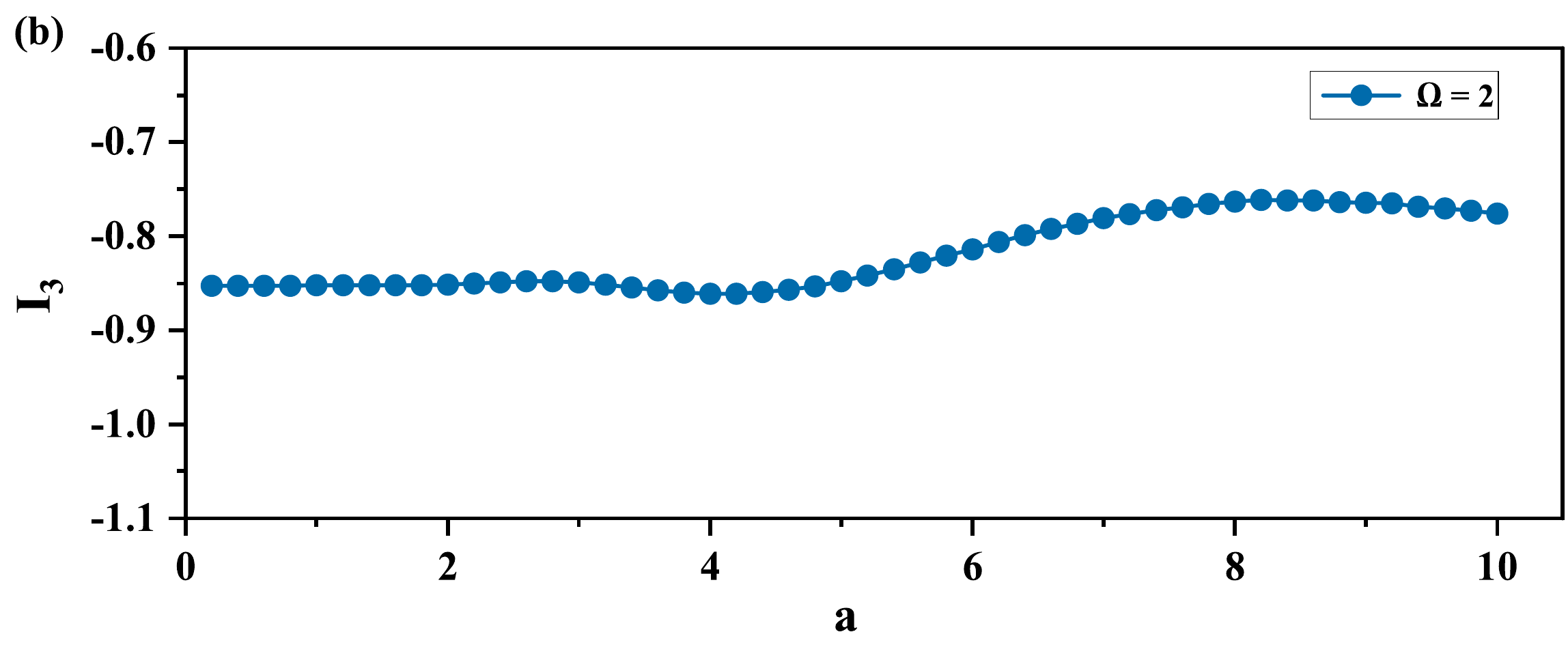}
\label{fig:2b}
\end{minipage}
\begin{minipage}[t]{0.5\textwidth}
\centering
\includegraphics[width=0.9\textwidth]{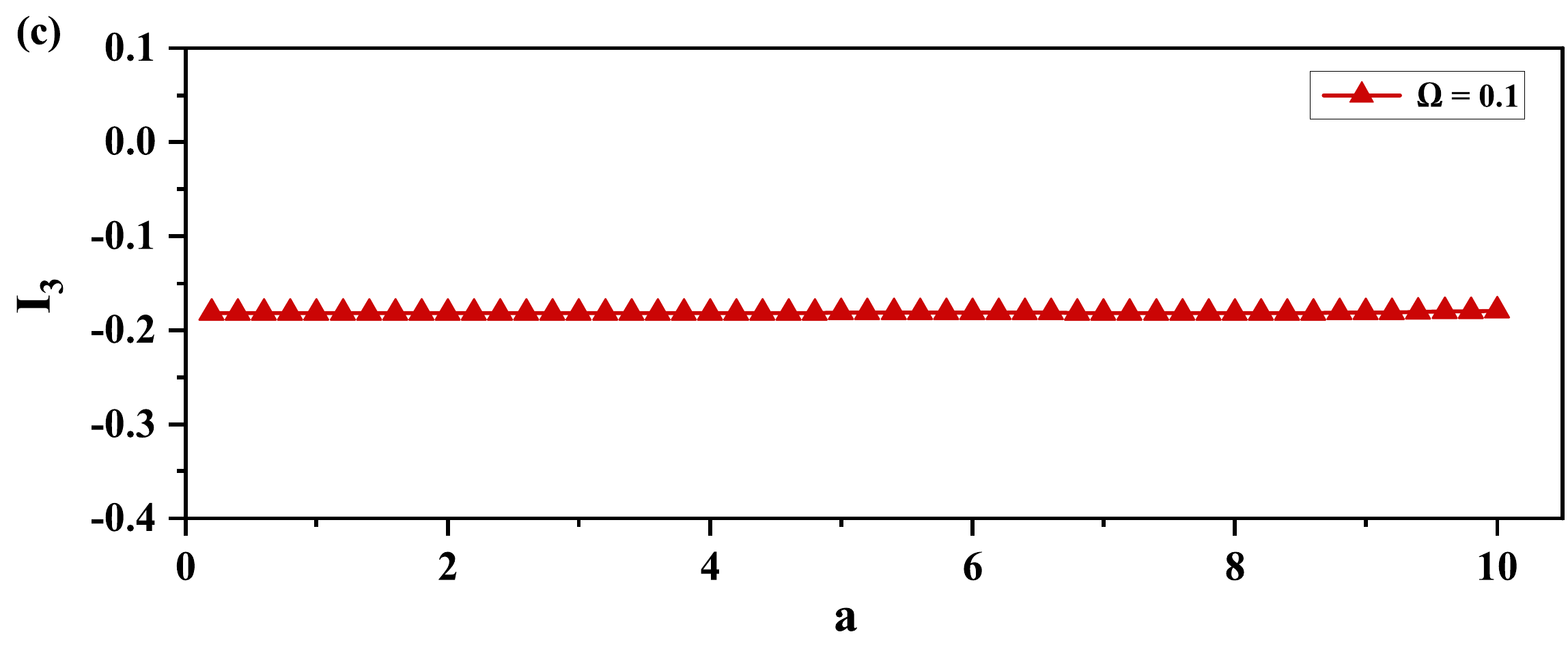}
\label{fig:2c}
\end{minipage}%
\begin{minipage}[t]{0.5\textwidth}
\centering
\includegraphics[width=0.9\textwidth]{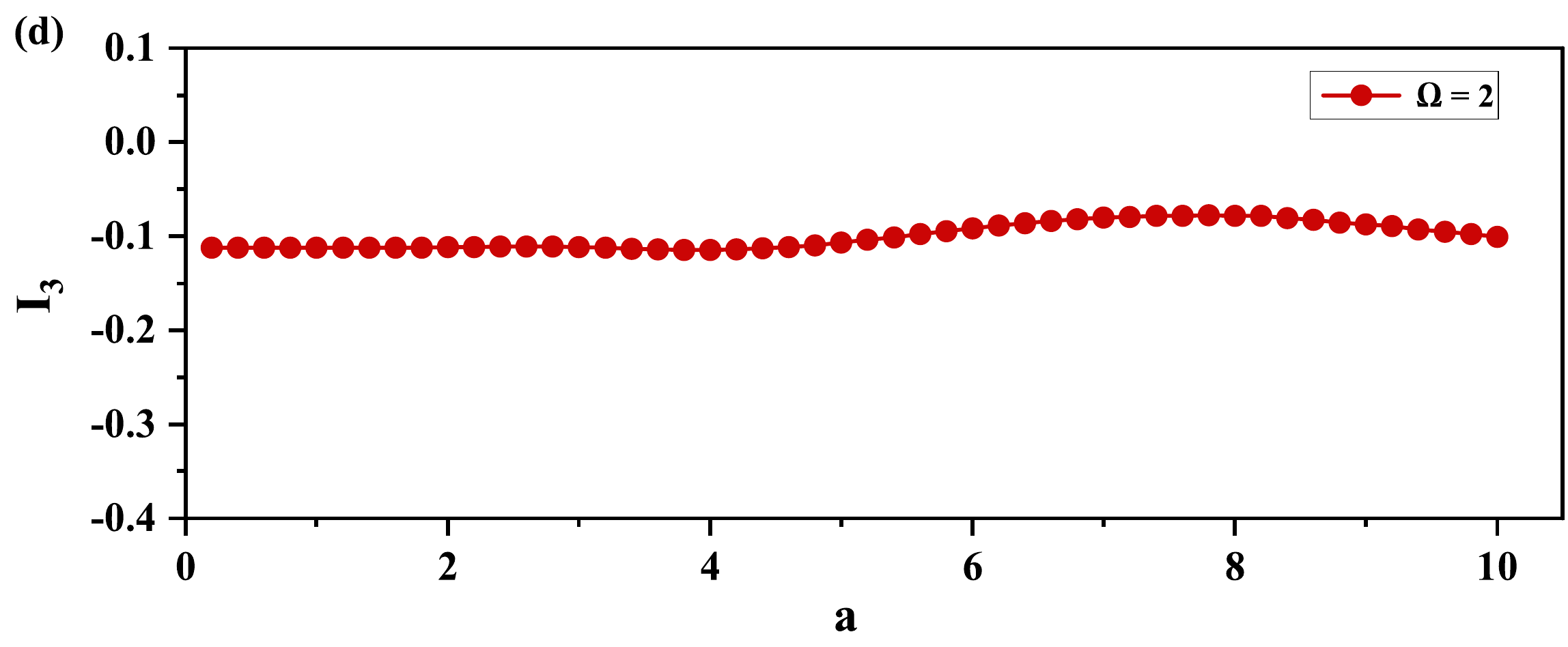}
\label{fig:2d}
\end{minipage}
\caption{The TMI as functions of the acceleration $a$ when Alice is in acceleration: (a) the initial state is GHZ state and $\Omega=0.1$; (b) the initial state is GHZ state and $\Omega=2$; (c) the initial state is W state and $\Omega=0.1$; (d) the initial state is W state and $\Omega=2$. The other parameters are the same as in Fig. 1.}
\end{figure*}

\subsection{Alice and Bob in acceleration}
Then we let Alice and Bob move with the same acceleration while Charlie remains stationary. 
In this case, the initial states evolve into 
\begin{equation}
\begin{aligned}
    \left|\psi\right\rangle_{A' B' C}^{GHZ}=& U_A U_B|\psi\rangle_{A B C}^{GHZ}\\
    = &\frac{1}{\sqrt{2}} [ C_0^2(\left|g_A g_B g_C 0_A 0_B 0_C\right\rangle- i \eta_0\left|g_A e_B g_C 0_A 1_B 0_C\right\rangle - i \eta_0\left|e_A g_B g_C 1_A 0_B 0_C\right\rangle \\
    &-  \eta_0^2 \left|e_A e_B g_C 1_A 1_B 0_C\right\rangle)+ C_1^2 (\left|e_A e_B e_C 0_A 0_B 0_C\right\rangle + i \eta_1\left|e_A g_B e_C 0_A 1_B 0_C\right\rangle \\
    &+ i \eta_1\left|g_A e_B e_C 1_A 0_B 0_C\right\rangle - \eta_1^2 \left|g_A g_B e_C 1_A 1_B 0_C\right\rangle )].
\end{aligned}
\end{equation}
and
\begin{equation}
\begin{aligned}
    \left|\psi\right\rangle_{A' B' C}^{W}=& U_A U_B|\psi\rangle_{A B C}^{W}\\
    = &\frac{1}{\sqrt{3}} [ C_0^2(\left|g_A g_B e_C 0_A 0_B 0_C\right\rangle- i \eta_0\left|g_A e_B e_C 0_A 1_B 0_C\right\rangle - i \eta_0\left|e_A g_B e_C 1_A 0_B 0_C\right\rangle \\
    &-  \eta_0^2 \left|e_A e_B e_C 1_A 1_B 0_C\right\rangle)+ C_0 C_1 (\left|g_A e_B g_C 0_A 0_B 0_C\right\rangle + i \eta_1\left|g_A g_B g_C 0_A 1_B 0_C\right\rangle \\
    &-  i \eta_0\left|e_A e_B g_C 1_A 0_B 0_C\right\rangle + \eta_0 \eta_1 \left|e_A g_B g_C 1_A 1_B 0_C\right\rangle + \left|e_A g_B g_C 0_A 0_B 0_C\right\rangle \\ 
    &- i \eta_0\left|e_A e_B g_C 0_A 1_B 0_C\right\rangle + i \eta_1\left|g_A g_B g_C 1_A 0_B 0_C\right\rangle + \eta_0 \eta_1 \left|g_A e_B g_C 1_A 1_B 0_C\right\rangle)].
\end{aligned}
\end{equation}

Adopting the approach used in the previous case, we obtain the expression for TMI and plot TMI ($I_3$) as a function of acceleration $a$ (refer to Fig. 3). 
There are noticeable similarities between Fig. 3 and Fig. 2. 
In Fig. 3(a), the TMI value is approximately -1.00, which is lower than the TMI values of about -0.7 in Fig. 3(b) and roughly -0.18 in Fig. 3(c). 
The same rules can be observed in other figures. 
The TMI value of Fig. 3(d) is about -0.1, which is larger than the TMI values of Fig. 3(b) and Fig. 3(c). 
As a result, we can draw a parallel conclusion to that of Fig. 2, namely, the anti-Unruh effect leads to greater information scrambling, and the W state is more stable than the GHZ state under identical conditions. 
\begin{figure*}
\begin{minipage}[t]{0.5\textwidth}
\centering
\includegraphics[width=0.9\textwidth]{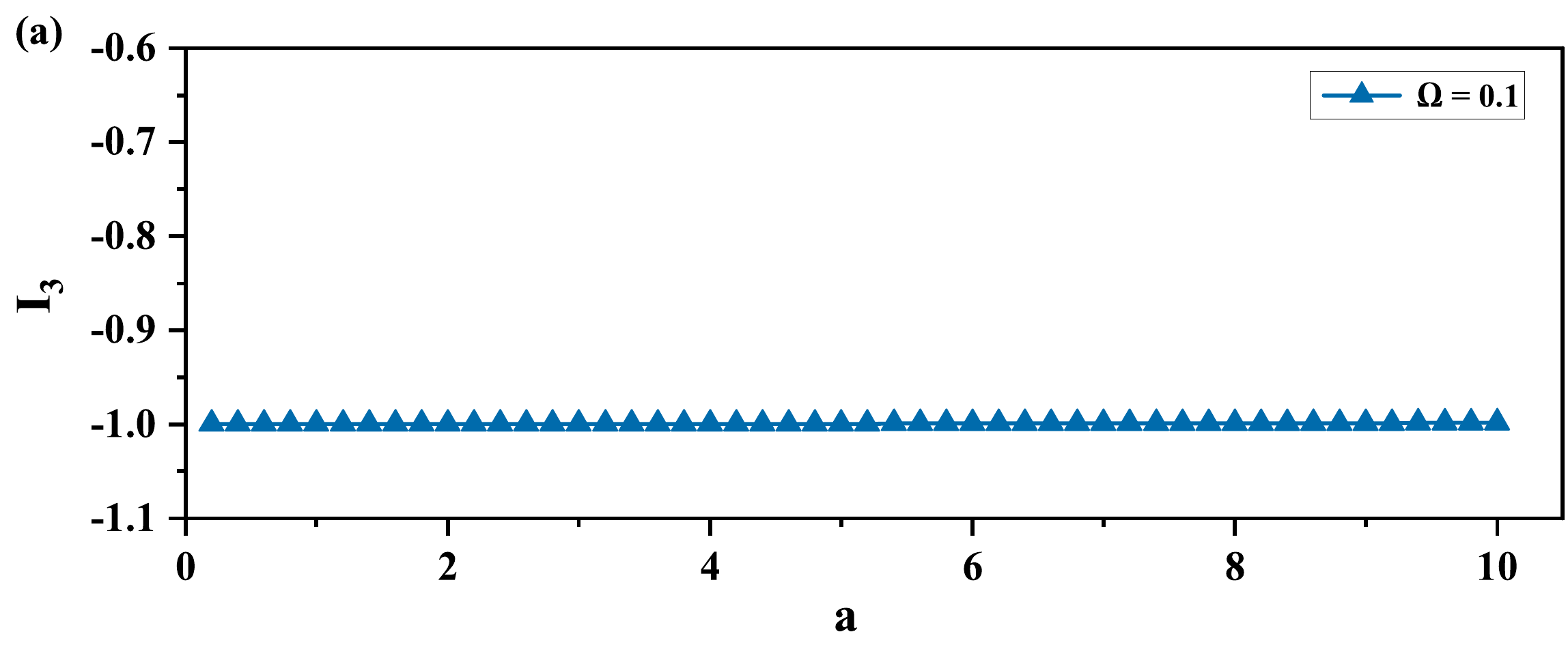}
\label{fig:3a}
\end{minipage}%
\begin{minipage}[t]{0.5\textwidth}
\centering
\includegraphics[width=0.9\textwidth]{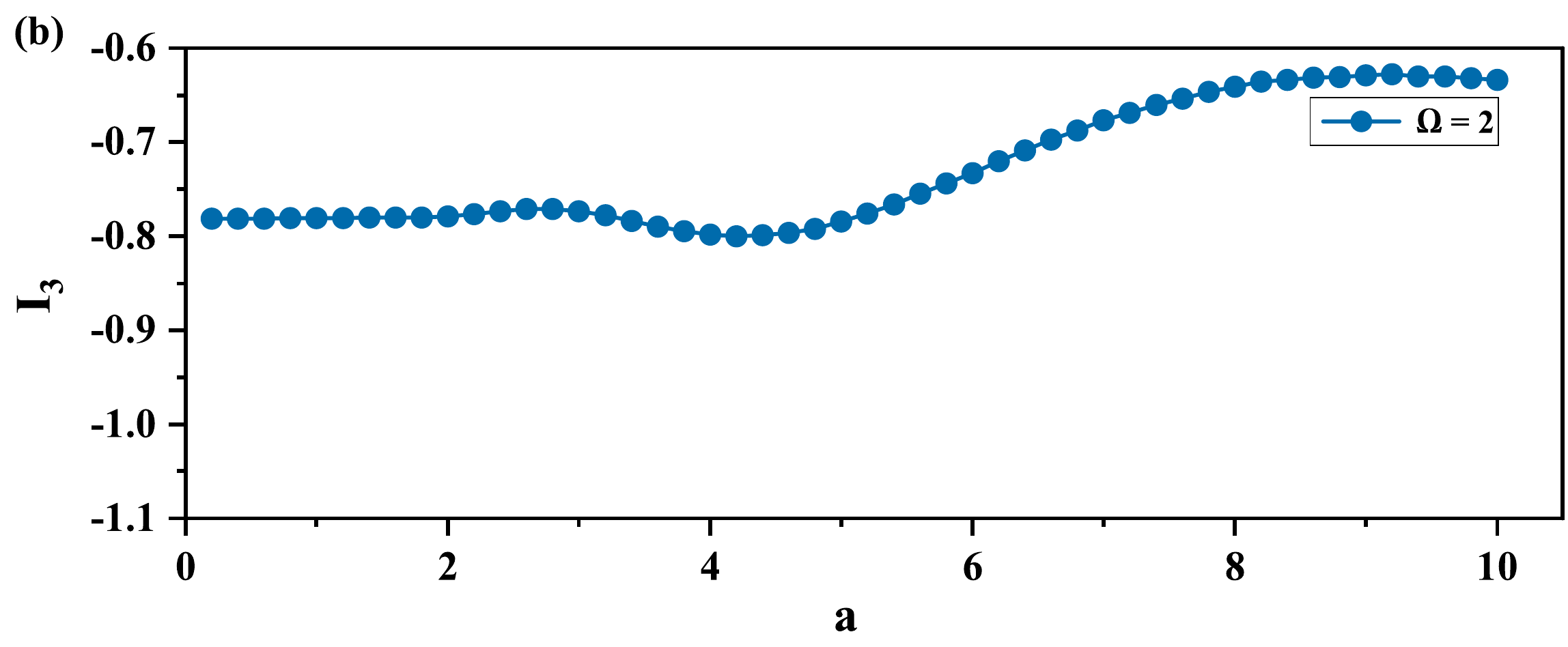}
\label{fig:3b}
\end{minipage}
\begin{minipage}[t]{0.5\textwidth}
\centering
\includegraphics[width=0.9\textwidth]{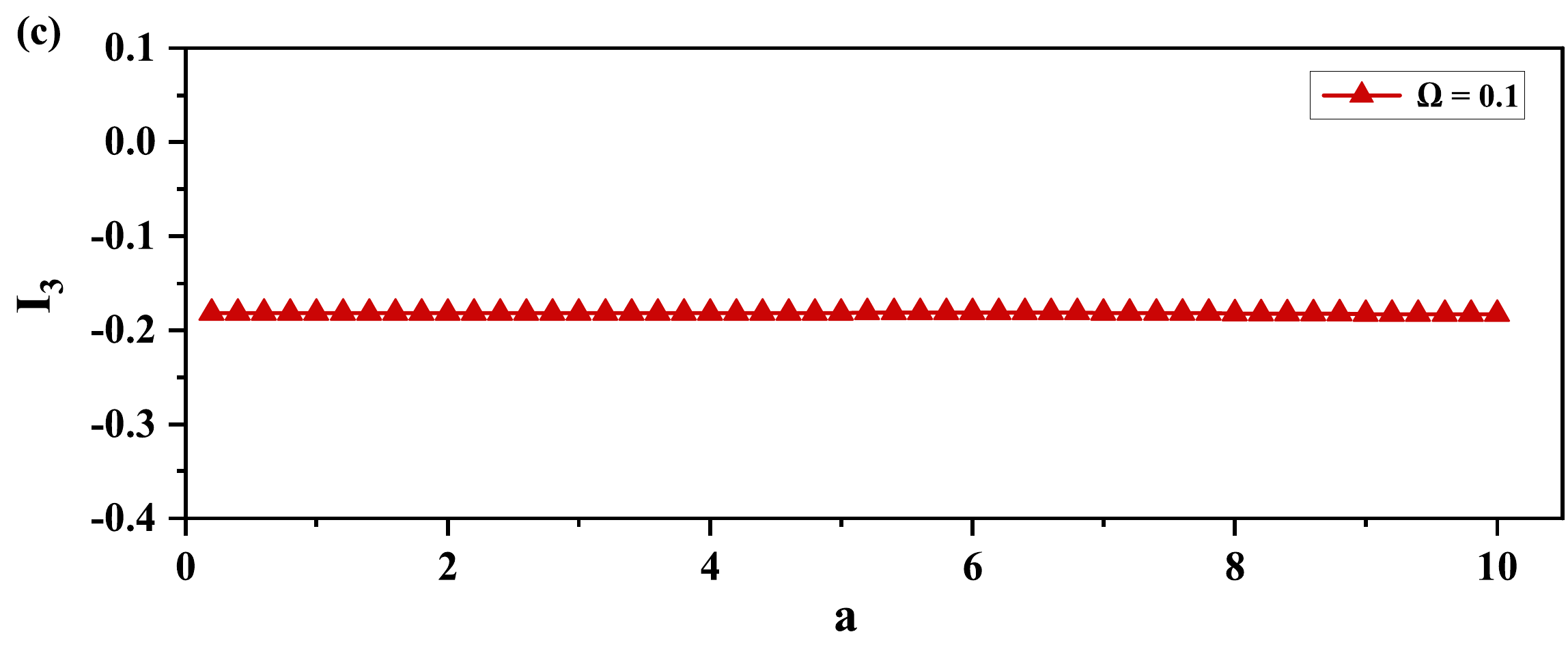}
\label{fig:3c}
\end{minipage}%
\begin{minipage}[t]{0.5\textwidth}
\centering
\includegraphics[width=0.9\textwidth]{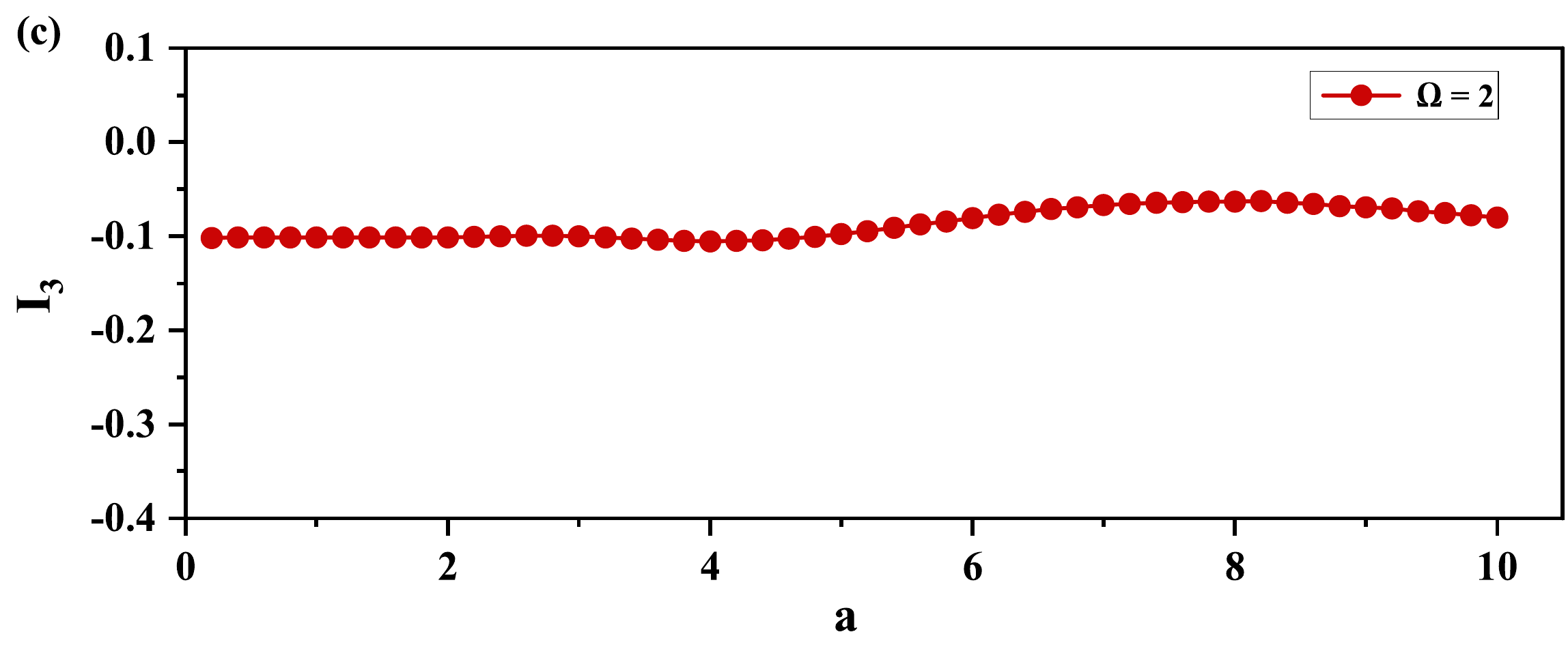}
\label{fig:3d}
\end{minipage}
\caption{The TMI as functions of the acceleration $a$ when Alice and Bob are in acceleration: (a) the initial state is GHZ state and $\Omega=0.1$; (b) the initial state is GHZ state and $\Omega=2$; (c) the initial state is W state and $\Omega=0.1$; (d) the initial state is W state and $\Omega=2$. The other parameters are the same as in Fig. 1.}
\end{figure*}

\subsection{All in acceleration}
Next, we examine the third situation that Alice, Bob, and Charlie all move with constant acceleration. 
By the transformation of Eq. (4), we can obtain 
\begin{equation}
\begin{aligned}
    \left|\psi\right\rangle_{A' B' C'}^{GHZ}=& U_A U_B U_C|\psi\rangle_{A B C}^{GHZ}\\
    =& \frac{1}{\sqrt{2}} [ C_0^3(\left|g_A g_B g_C 0_A 0_B 0_C\right\rangle- i \eta_0\left|g_A g_B e_C 0_A 0_B 1_C\right\rangle - i \eta_0\left|g_A e_B g_C 0_A 1_B 0_C\right\rangle \\
    &-  \eta_0^2 \left|g_A e_B e_C 0_A 1_B 1_C\right\rangle - i\eta_0 \left|e_A g_B g_C 1_A 0_B 0_C\right\rangle- \eta_0^2 \left|e_A g_B e_C 1_A 0_B 1_C\right\rangle \\
    &-  \eta_0^2 \left|e_A e_B g_C 1_A 1_B 0_C\right\rangle + i \eta_0^3 \left|e_A e_B e_C 1_A 1_B 1_C\right\rangle)+ 
    C_1^3 (\left|e_A e_B e_C 0_A 0_B 0_C\right\rangle\\
    &+ i \eta_1\left|e_A e_B g_C 0_A 0_B 1_C\right\rangle + i \eta_1\left|e_A g_B e_C 0_A 1_B 0_C\right\rangle 
    - \eta_1^2 \left|e_A g_B g_C 0_A 1_B 1_C\right\rangle \\
    &+  i\eta_1 \left|g_A e_B e_C 1_A 0_B 0_C\right\rangle- \eta_1^2 \left|g_A e_B g_C 1_A 0_B 1_C\right\rangle 
    - \eta_1^2 \left|g_A g_B e_C 1_A 1_B 0_C\right\rangle \\
    &- i \eta_1^3 \left|g_A g_B g_C 1_A 1_B 1_C\right\rangle)],
\end{aligned}
\end{equation}
and
\begin{equation}
\begin{aligned}
    \left|\psi\right\rangle_{A' B' C'}^{W}=& U_A U_B U_C|\psi\rangle_{A B C}^{W}\\
     =& \frac{1}{\sqrt{3}} C_0^2 C_1 (\left|g_A g_B e_C 0_A 0_B 0_C\right\rangle + i \eta_1\left|g_A g_B g_C 0_A 0_B 1_C\right\rangle - i \eta_0\left|g_A e_B e_C 0_A 1_B 0_C\right\rangle \\
    &+ \eta_0 \eta_1 \left|g_A e_B g_C 0_A 1_B 1_C\right\rangle - i\eta_0 \left|e_A g_B e_C 1_A 0_B 0_C\right\rangle + \eta_0 \eta_1 \left|e_A g_B g_C 1_A 0_B 1_C\right\rangle \\
    &- \eta_0^2 \left|e_A e_B e_C 1_A 1_B 0_C\right\rangle - i \eta_0^2 \eta_1 \left|e_A e_B g_C 1_A 1_B 1_C\right\rangle+
    \left|g_A e_B g_C 0_A 0_B 0_C\right\rangle\\
    &- i \eta_0\left|g_A e_B e_C 0_A 0_B 1_C\right\rangle + i \eta_1\left|g_A g_B g_C 0_A 1_B 0_C\right\rangle 
    + \eta_0 \eta_1 \left|g_A g_B e_C 0_A 1_B 1_C\right\rangle \\
    &- i\eta_0 \left|e_A e_B g_C 1_A 0_B 0_C\right\rangle- \eta_0^2 \left|e_A e_B e_C 1_A 0_B 1_C\right\rangle 
    + \eta_0\eta_1 \left|e_A g_B g_C 1_A 1_B 0_C\right\rangle \\
    &- i \eta_0^2\eta_1 \left|e_A g_B e_C 1_A 1_B 1_C\right\rangle +  \left|e_A g_B g_C 0_A 0_B 0_C\right\rangle
    -  i \eta_0\left|e_A g_B e_C 0_A 0_B 1_C\right\rangle \\
    &- i \eta_0\left|e_A e_B g_C 0_A 1_B 0_C\right\rangle - \eta_0^2 \left|e_A e_B e_C 0_A 1_B 1_C\right\rangle 
    +  i\eta_1 \left|g_A g_B g_C 1_A 0_B 0_C\right\rangle\\
    &+ \eta_0 \eta_1 \left|g_A g_B e_C 1_A 0_B 1_C\right\rangle + \eta_0 \eta_1 \left|g_A e_B g_C 1_A 1_B 0_C\right\rangle 
    - i \eta_0^2 \eta_1 \left|g_A e_B e_C 1_A 1_B 1_C\right\rangle).
\end{aligned}
\end{equation}

Similarly, we can plot TMI ($I_3$) as a function of acceleration $a$ (refer to Fig. 4). 
The TMI value in Fig. 4(a) is approximately -2.75, which is lower than the TMI value in Fig. 4(b). 
Notably, the TMI values in Fig. 4(c) and Fig. 4(d) are positive, indicating the absence of information scrambling under the corresponding conditions. 
The observed phenomena suggest that the anti-Unruh effect contributes to stronger information scrambling throughout the system, while the W state results in weaker or even nonexistent information scrambling. 
This finding is in line with the outcomes of previous cases.
\begin{figure*}
\begin{minipage}[t]{0.5\textwidth}
\centering
\includegraphics[width=0.9\textwidth]{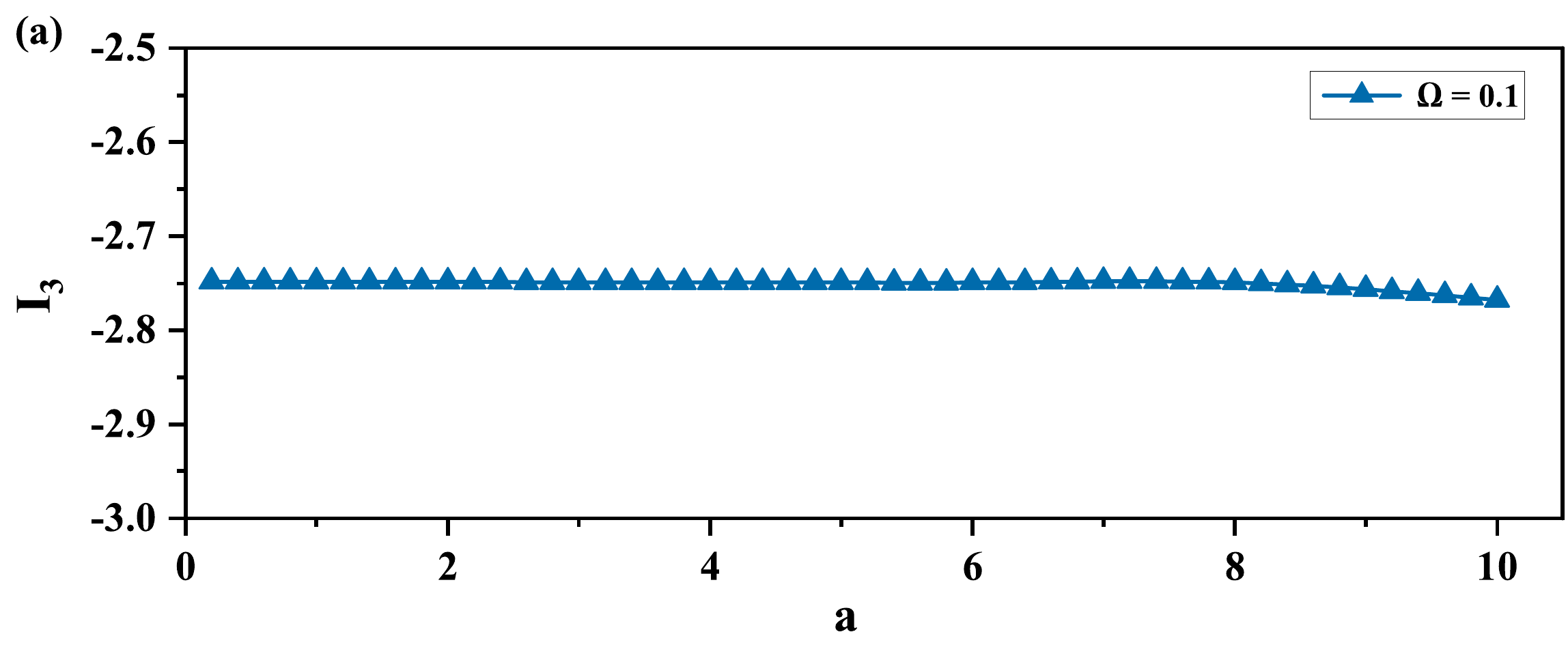}
\label{fig:4a}
\end{minipage}%
\begin{minipage}[t]{0.5\textwidth}
\centering
\includegraphics[width=0.9\textwidth]{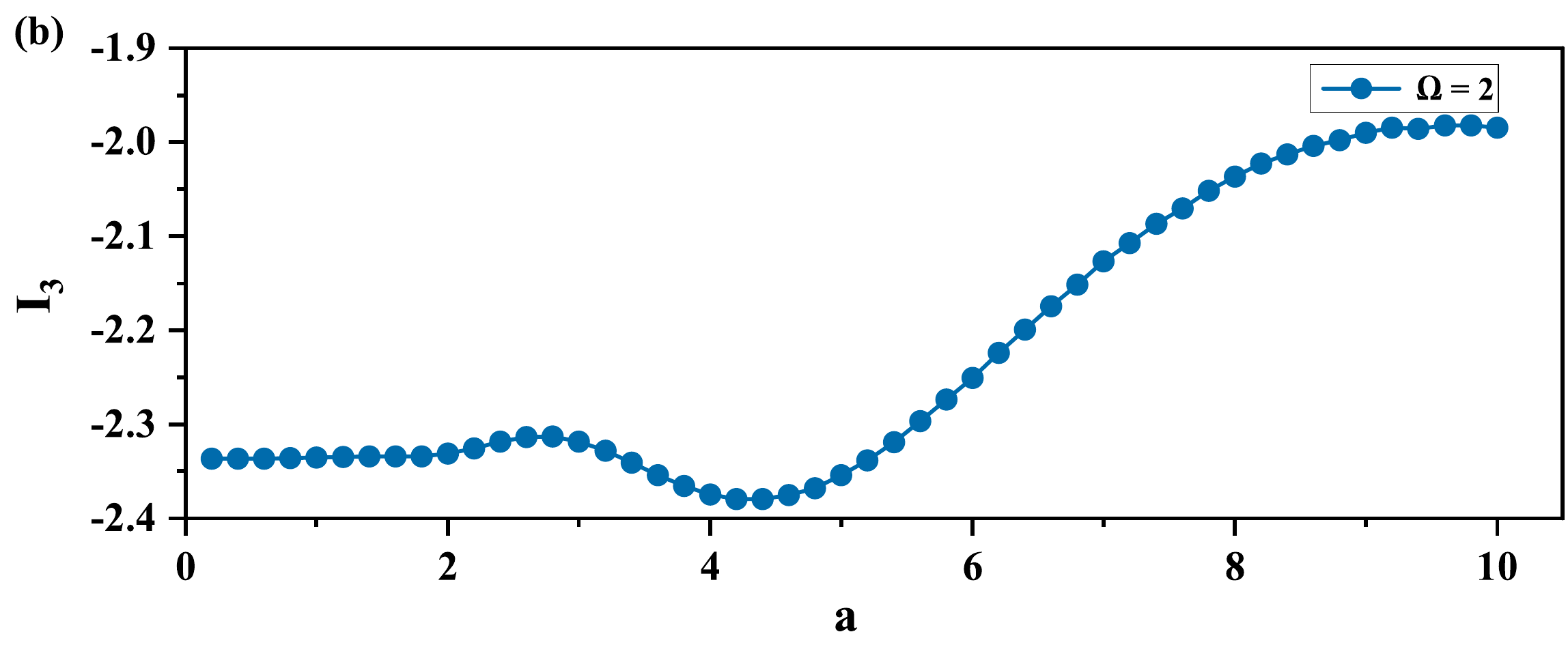}
\label{fig:4b}
\end{minipage}
\begin{minipage}[t]{0.5\textwidth}
\centering
\includegraphics[width=0.9\textwidth]{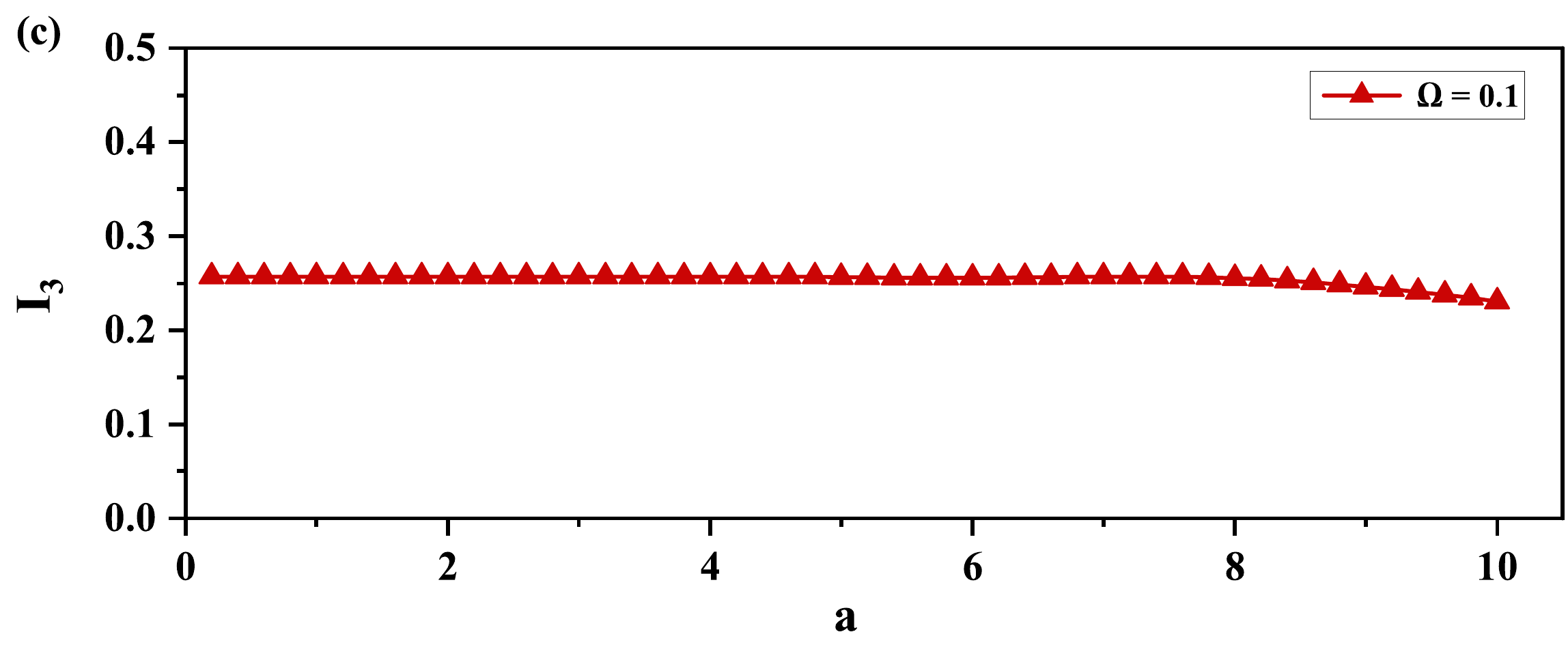}
\label{fig:4c}
\end{minipage}%
\begin{minipage}[t]{0.5\textwidth}
\centering
\includegraphics[width=0.9\textwidth]{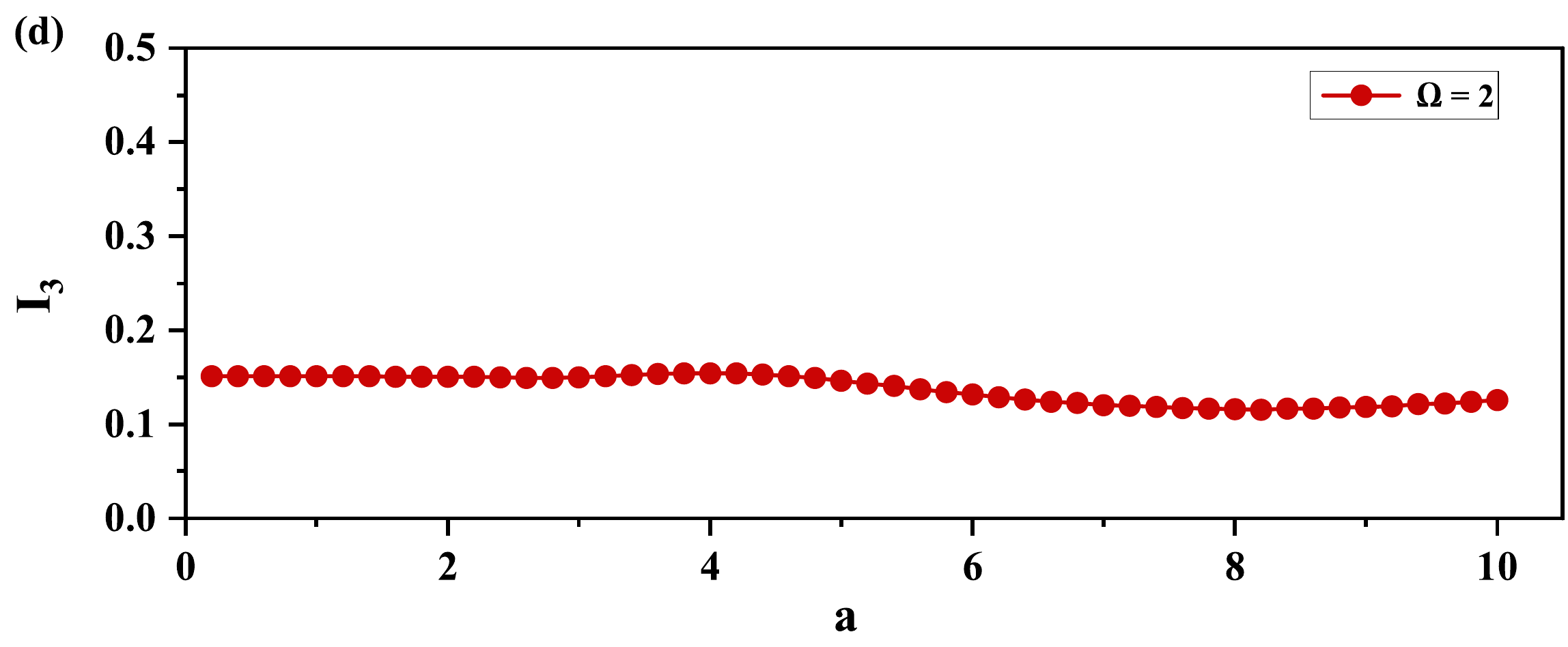}
\label{fig:4d}
\end{minipage}
\caption{The TMI as functions of the acceleration $a$ when All are in acceleration: (a) the initial state is GHZ state and $\Omega=0.1$; (b) the initial state is GHZ state and $\Omega=2$; (c) the initial state is W state and $\Omega=0.1$; (d) the initial state is W state and $\Omega=2$. The other parameters are the same as in Fig. 1.}
\end{figure*}

\section{The case of multiple detectors}
\label{sec:n}
For the $N$ detectors case, the calculation of TMI is much more complicated. 
Here, we only discuss the $N$-partite GHZ state and $N$-partite product state. 
\subsection{$N$-partite GHZ state}
We consider the $N$-partite GHZ state, 
\begin{equation}
    \left|\psi\right\rangle_{1,2,...,N}^{GHZ}=\frac{1}{\sqrt{2}} (\left|g_1 g_2... g_N\right\rangle + \left|e_1 e_2... e_N\right\rangle)\left|0_1 0_2... 0_N\right\rangle.
\end{equation}
Assume the first and second detectors are in acceleration, 
\begin{equation}
\begin{aligned}
    \left|\psi\right\rangle_{1',2',...,N}^{GHZ}=& U_1 U_2\left|\psi\right\rangle_{1,2,...,N}^{GHZ}\\
    = &\frac{1}{\sqrt{2}} [ C_0^2(\left|g_1 g_2... g_N 0_1 0_2... 0_N\right\rangle- i \eta_0\left|g_1 e_2... g_N 0_1 1_2... 0_N\right\rangle - i \eta_0\left|e_1 g_2... g_N 1_1 0_2... 0_N\right\rangle \\
    &-  \eta_0^2 \left|e_1 e_2... g_N 1_1 1_2... 0_N\right\rangle)+ C_1^2 (\left|e_1 e_2... e_N 0_1 0_2... 0_N\right\rangle + i \eta_1\left|e_1 g_2... e_N 0_1 1_2... 0_N\right\rangle \\
    &+ i \eta_1\left|g_1 e_2... e_N 1_1 0_2... 0_N\right\rangle - \eta_1^2 \left|g_1 g_2... e_N 1_1 1_2... 0_N\right\rangle )].
\end{aligned}
\end{equation}

Now, we can employ two distinct methods to partition the $N$-partite system into three subsystems: A, B, and C. 
The first method involves assigning the first detector to subsystem A, the second detector to subsystem B, and the remaining detectors to subsystem C. 
The second method involves designating the first and second detectors as subsystem A, dividing the remaining detectors equally between subsystems B and C. 
For the first method, we calculate that the TMI expression is identical to the case of Alice and Bob accelerating in three detectors, as depicted in Fig. 3(a) and Fig. 3(b). 
For the second method, we can derive the TMI 
\begin{equation}
\begin{aligned}
 I_3= &\log{2} - \frac{1}{2}(C_0^4 + C_1^4)\log{[\frac{1}{2}(C_0^4 + C_1^4)]} 
 - \frac{1}{2}(C_1^4 + C_0^4\eta_0^4)\log{[\frac{1}{2}(C_1^4 + C_0^4\eta_0^4)]}\\ 
 &+ C_0^4\log{(\frac{1}{2}C_0^4)} + C_1^4\log{(\frac{1}{2}C_1^4)}
 +2 C_0^4\eta_0^2\log{(\frac{1}{2}C_0^4\eta_0^2)} + C_0^4\eta_0^4\log{(\frac{1}{2}C_0^4\eta_0^4)}\\ 
 &+ 2 C_1^4\eta_1^2\log{(\frac{1}{2}C_1^4\eta_1^2)} + C_1^4\eta_1^4\log{(\frac{1}{2}C_1^4\eta_1^4)} 
 - 2(C_0^4\eta_0^2 + C_1^4\eta_1^2)\log{[\frac{1}{2}(C_0^4\eta_0^2 + C_1^4\eta_1^2)]}\\
 &- \frac{1}{2}(C_0^4 + C_1^4\eta_1^4)\log{[\frac{1}{2}(C_0^4 + C_1^4\eta_1^4)]} 
 - \frac{1}{2}(C_0^4\eta_0^4 + C_1^4\eta_1^4)\log{[\frac{1}{2}(C_0^4\eta_0^4 + C_1^4\eta_1^4)]}, 
\end{aligned}
\end{equation}
and obtain the TMI$(I_3)$ as functions of the acceleration $a$, as illustrated in Fig. 5. 
It is easy to observe that the TMI value is essentially around -1.00 in Fig. 5(a), which is lower than the value of approximately -0.65 in Fig. 5(b). 
Consequently, we can conclude that the anti-Unruh effect induces greater information scrambling, which is consistent with the findings in the previous sections.
\begin{figure}
\begin{minipage}[t]{0.5\textwidth}
\centering
\includegraphics[width=0.9\textwidth]{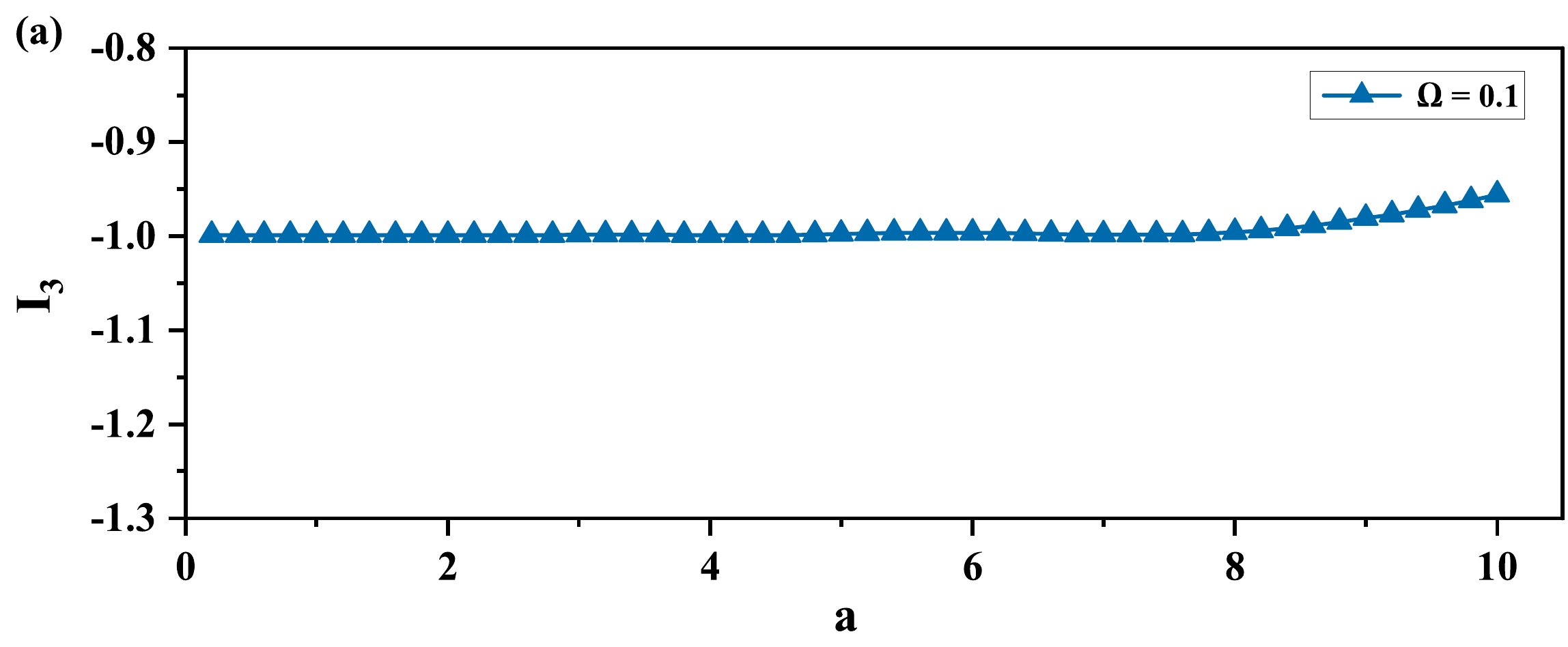}
\label{fig:5a}
\end{minipage}%
\begin{minipage}[t]{0.5\textwidth}
\centering
\includegraphics[width=0.9\textwidth]{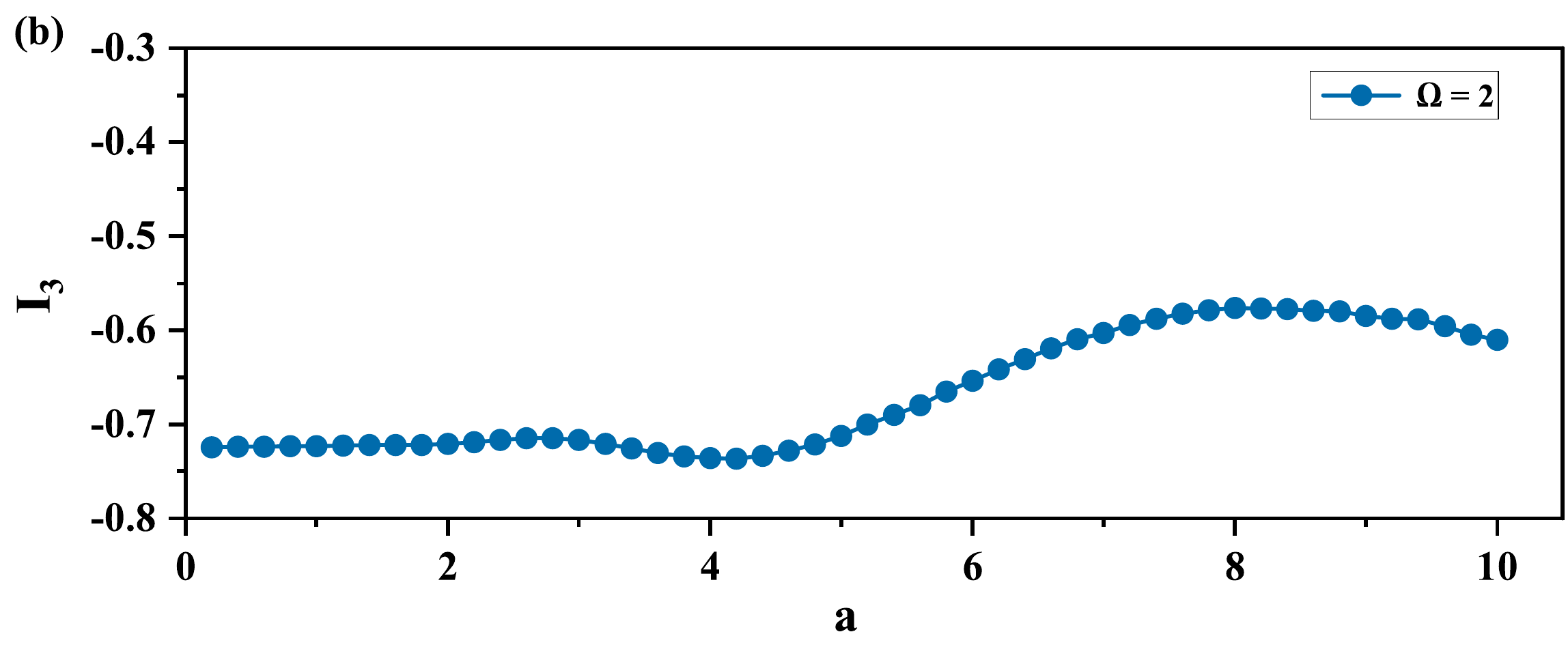}
\label{fig:5b}
\end{minipage}
\caption{The TMI as functions of the acceleration $a$ in $N$ detectors case when the first and second detectors are in acceleration: (a) the initial state is GHZ state and $\Omega=0.1$; (b) the initial state is GHZ state and $\Omega=2$. The other parameters are the same as in Fig. 1.}
\label{fig:5}
\end{figure}

\subsection{$N$-partite product state}
In the preceding section, we presented the evolution of TMI when the detectors are in entangled states. 
Nevertheless, there may initially be no correlation between the three detectors. 
Assuming the initial state of detectors is a product state, such as $\left|\psi\right\rangle_{1, 2,..., N}=\left|e g... g\right\rangle\left|0 0... 0\right\rangle$, if the first detector is in acceleration, the initial product state will transform into:
\begin{equation}
\begin{aligned}
    \left|\psi\right\rangle_{1', 2,..., N}=& U_1 |\psi\rangle_{1, 2,..., N}\\
    = & C_1\left|e_1 g_2... g_N 0_1 0_2... 0_N\right\rangle+i C_1 \eta_1\left|g_1 g_2... g_N 1_1 0_2... 0_N\right\rangle.
\end{aligned}
\end{equation}
Tracing over the scalar field modes, the reduced density matrix $\rho_{A' B C}$ can be expressed as:
\begin{equation}
\begin{aligned}
    \rho_{1', 2,..., N}= & C_1^2(\left|e_1 g_2... g_N \right\rangle \left\langle e_1 g_2... g_N\right| + \eta_1^2\left|g_1 g_2... g_N \right\rangle\left\langle g_1 g_2... g_N\right|)\\
    = & C_1^2 (\left|e_1 \right\rangle \left\langle e_1\right|+ \eta_1^2\left|g_1 \right\rangle\left\langle g_1 \right|)\otimes \left| g_2... g_N \right\rangle \left\langle  g_2... g_N\right|\\
    = & \rho_{1'} \otimes \rho_{2}\otimes...\otimes \rho_{N}. 
\end{aligned}
\end{equation}

Moreover, it is straightforward to determine that the TMI remains zero. 
By simply recalculating, we can derive a similar result when two or more detectors are in acceleration. 
In other words, information scrambling does not occur when the detectors are initially in a product state.

\section{Discussion and conclusion}
\label{sec:conclusion}
In summary, we employed the Unruh-DeWitt model to systematically investigate the impact of acceleration on information scrambling quantified by TMI in cases involving the Unruh effect and the anti-Unruh effect. 
Our numerical calculations reveal that stronger information scrambling occurs under the anti-Unruh effect. 
In addition, we found that the GHZ state is more susceptible to acceleration and generates greater information scrambling than the W state. 
We also extended our analysis to the $N$-detector case, i.e., the $N$-partite GHZ state, and found results consistent with the tripartite system. 
Finally, we demonstrated that information is not scrambled when the detectors are initially in a product state. 
It is worth noting that recent studies have identified a phenomenon similar to the anti-Unruh effect in certain physical situations, known as the anti-Hawking effect~\cite{henderson2020anti, robbins2022anti}. 
Further exploration using the methods presented in this paper can also contribute to our understanding of the Hawking effect and the anti-Hawking effect.

\acknowledgments

The author thanks Qing-yu Cai and Baocheng Zhang for valuable discussions and comments.
This work was supported by the National Natural Science Foundation of China under Grant No. 11725524.

% Bibliography

%% [A] Recommended: using JHEP.bst file
%% \bibliographystyle{JHEP}
%% \bibliography{biblio.bib}

%% or
%% [B] Manual formatting (see below)
%% (i) We suggest to always provide author, title and journal data or doi:
%% in short all the informations that clearly identify a document.
%% (ii) please avoid comments such as "For a review'', "For some examples",
%% "and references therein" or move them in the text. In general, please leave only references in the bibliography and move all
%% accessory text in footnotes.
%% (iii) Also, please have only one work for each \bibitem.

\end{document}